# Analysis of charge transfer mechanism on $(Ba_{1-x}Nd_xCuO_{2+\delta})_2/(CaCuO_2)_n$ superconducting superlattices by thermoelectric power measurements

M.Putti, D.Marré, I.Pallecchi,

*INFM-LAMIA, Dipartimento di Fisica, Via Dodecaneso 33, 16146 Genova, Italy*

P.G. Medaglia, A. Tebano, G. Balestrino

*INFM-COHERENTIA, Dipartimento di Ingegneria Meccanica, Università d Roma Tor Vergata,*

*Via del Politecnico 1, 00133 Roma*

**ABSTRACT**

We have investigated the charge transfer mechanism in artificial superlattices by Seebeck effect measurements. Such a technique allows a precise determination of the amount of charge transferred on each $CuO_2$ plane. A systematic characterization of thermoelectric power in $(BaCuO_{2+\delta})_2/(CaCuO_2)_n$ and $(Ba_{0.9}Nd_{0.1}CuO_{2+\delta})_2/(CaCuO_2)_n$ superlattices demonstrates that electrical charge distributes uniformly among the $CuO_2$ planes in the Ca-block. The differences observed in the Seebeck effect behavior between the Nd-doped and undoped superlattices are ascribed to the different metallic character of the Ba-block in the two cases. Finally, the special role of structural disorder in superlattices with *n=1* is pointed out by such analysis.

**Introduction**

The crystallographic structure of the infinite layers (IL) compounds $ACuO_2$ (A= alkaline earth) consists of $CuO_2$ planes, considered the essential feature for high temperature superconductivity (HTS), separated by alkaline earth planes. Undoped IL compounds are non-superconducting. IL-based superlattices are formed by two different compounds alternately stacked in sequence: the first layer is either $CaCuO_2$ or $SrCuO_2$, the second is $BaCuO_2$. The Ba IL compound is very unstable and tends to incorporate apical extra oxygen ions giving rise to a different phase with a doubled c lattice parameter (1:2:0:1 phase). Thus, $CaCuO_2/BaCuO_{2+\delta}$ superlattices, grown at relatively high oxygen pressure ($P_{O2}\approx 1$ mbar), contain extra oxygen ions in the Ba-block. The essential feature of the Ba-block ("charge reservoir" or CR block) is then a sizeable electrical charge unbalance. Under these conditions holes are transferred from the CR block into the $CaCuO_2$ block. As a consequence the Ca-block becomes superconducting [1]. Superconducting properties of the material can be engineered varying the structure of the two constituent blocks. Following this approach electrical transport properties of $(BaCuO_{2+\delta})_2/(CaCuO_2)_n$ superlattices have been investigated as a function of the number of superconducting $CuO_2$ planes, by varying the thickness of the IL bock. The number $n$ of $CaCuO_2$ layers has been varied between 1 and 15, while keeping constant all the remaining parameters, including the CR block thickness, so as to change the doping level per $CuO_2$ plane inside the IL superconducting block [2].

The behavior of the critical temperature $T_c$ versus $n$ shows a number of interesting features:

1. the maximum value of $T_c$ is reached for $n=2-3$;
2. a steep decrease of $T_c$ occurs for $n=1$;
3. for $n>3$ $T_c$ decreases gradually until reaching an insulating state for $n>11$.

The strong reduction of the transition temperature in the $n=1$ superlattice has been tentatively ascribed to a reduction of the quasiparticle mean free path caused by the disorder at the interfaces. The $T_c$ decrease for $n>3$ has been explained by taking into account the decrease of the effective doping $p_{pl}$ (number of holes per number of $CuO_2$ plane in unit cell) of the IL block which occurs when the thickness of the IL block increases. However, no direct evidence has been given so far of such a doping mechanism.

In this paper we investigate the actual doping of the $CuO_2$ planes of superlattices with different periodicity and/or chemical composition, by means of Seebeck effect measurements, which represent an unique tool for the analysis of the carrier concentration in HTS compounds.

All the electronic properties of HTS materials show a close relationship with the charge carrier concentration of the $CuO_2$ planes. Among them, an impressive example is the Seebeck effect. Namely, the Seebeck coefficient S shows an absolute value and, moreover, an unusual temperature dependence which depend on $p_{pl}$ in a very general way, almost independent on the specific HTS compound investigated [3-5]. In particular, the absolute value of $S(T)$ exhibits a strong systematic decrease with $p_{pl}$. The room temperature value of the Seebeck effect taken as a convenient parameter, falls nearly exponentially with $p_{pl}$ from 500 μV/K, for the underdoped samples ($0<p_{pl}<0.16$), towards 1-2 μV/K in the case of optimal doping ($p_{pl} \approx 0.16$), and, finally, reaches the negative value -15 μV/K for $p_{pl} \approx 0.3$ (maximum doping allowed for high temperature superconductivity). Despite some deviations found in YBCO compounds, where fully oxygenated CuO chains become metallic and give a negative contribution to the Seebeck effect [5], most cuprates follow closely this general trend so that an universal relation between the Seebeck effect at 290 K and $p_{pl}$ was established by Orbetelli, Cooper and Tallon (OCT) and conveniently parameterized by the relations [4]:

$$S^{OCT} = 372 \exp(-32.4 p_{pl}) \text{ μV/K} \qquad p_{pl}<0.05$$

$$S^{OCT} = 992 \exp(-38.1 p_{pl}) \quad \mu V/K \qquad 0.05 < p_{pl} < 0.15 \qquad (1)$$

$$S^{OCT} = -139 p_{pl} + 24.2 \quad \mu V/K \qquad p_{pl} > 0.155$$

These relations provide, in principle, a quantitative evaluation of the doping of $CuO_2$ planes.

The OCT law was applied to many cuprate supercoductors, [3-10] and superlattices [11,12], and, apart from few cases [7,8], this law seems to account for both single- and multiple- $CuO_2$ layers superconductors, despite of their large differences in structure and chemical composition. The applicability of OCT law to many compounds in different forms (polycrystals, thin films and single crystals) is based on two important features of the Seebeck effect: it is not affected by granularity [13] and, within some extents, it is independent on impurity scattering rates. In this respect Seebeck effect differs from electrical resistivity which strongly depends on granularity, disorder and localization effects which mask its dependence on carrier concentration. Indeed, the OCT law has been successfully applied to polycrystals and thin films, with the only caution of taking the Seebeck effect anisotropy into account; in particular, eq.(1) describes the Seebeck coefficient of poly-crystals and can be easily extended to the *ab*-plane Seebeck coefficient measured in thin films which is 3/2 times as much. Thus, thermopower measurements have proved to be a precious tool to evaluate quantitatively the $CuO_2$ planes doping in novel cuprates, substituted compounds and artificial superlattices.

In this paper we present a systematic analysis of the thermopower in artificial $(BaCuO_{2+\delta})_2/(CaCuO_2)_n$ and $(Ba_{0.9}Nd_{0.1}CuO_{2+\delta})_2/(CaCuO_2)_n$ superlattices. Our goal is to extract a precise evaluation of the doping level of the $CuO_2$ planes which is of crucial importance in order to explain on a quantitative basis the behavior of $T_c$ versus $n$.

**Experimental**

$(BaCuO_{2+\delta})_2/(CaCuO_2)_n$ and $(Ba_{0.9}Nd_{0.1}CuO_{2+\delta})_2/(CaCuO_2)_n$ superconducting superlattices were grown by Pulsed Laser Deposition (PLD) in high oxygen pressure ($P_{O2} \approx 1$ mbar) at a substrate temperature of about 600 °C. Trivalent ion (Nd) substitutions in the Ba site were introduced so as to reduce the charge unbalance in the CR block due to excess oxygen, thus improving its chemical stability [14]. A KrF excimer laser ($\lambda$=248 nm) with a pulse length of 25 nm and a pulse energy of 300 mJ was used for the growth. In our experimental set-up the laser beam forms an angle of 45° relative to the target surface and the beam spot is focused to 3 mm$^2$ at the target surface. Targets were mounted on a computer controlled carousel. The number of laser shots on each target was adjusted in order to grow different artificial structures. During deposition each target rotated around the perpendicular to the surface. $BaCuO_{2+\delta}$, $Ba_{0.9}Nd_{0.1}CuO_{2+\delta}$ and $CaCuO_2$ targets were prepared according the following procedure: stoichiometric mixtures of high purity $CaCO_3$, $BaCO_3$, $CuO$ and $Nd_2O_3$ powders were calcinated at 860°C in air for at least 24 hours. The weight loss was verified. Powders were then pressed in a disk shape and then sintered for 24 hours at 900 °C. Pellets resulting from this procedure were not single phase. $SrTiO_3$ ((001) oriented) substrates were placed on a heated holder at a distance of 5 cm from the target. After deposition films were quenched to room temperature in few minutes in an oxygen atmosphere. Structural properties of films were investigated by X-ray diffraction using a $\Theta$-$2\Theta$ Bragg-Brentano diffractometer. Transport properties were measured by the standard four probes technique. Silver contacts were deposited on the substrate surface prior to film growth in order to decrease the contact resistance. At the end of the deposition procedure, an amorphous protecting layer of electrically insulating $CaCuO_2$ was deposited on the top of the film (deposition temperature lower than 100 C°).

The Seebeck effect measurements were performed in a home-built cryostat working from 10 to 300 K. The thermopower was measured in a steady flux configuration in which the heat flow is supplied to one end of the sample by a strain-gauge heater ($1\times1mm^2$), while the other end is in thermal contact with the heat sink. The temperature difference was measured with Au(Fe)-Kp thermocouples, and the electrical contacts to 99.99% Cu wires were made with Ag varnish painted on the Ag contacts previously deposited on the substrate. To avoid spurious thermopowers we use an a.c. technique described elsewhere [15]. The gradient applied to the sample was varied from 1 to 0.1 K/cm; the frequency was chosen as low as $\nu=0.005-0.003$ Hz in order to avoid a reduction of the heat wave amplitude along the sample.

**Results and discussion**

As a preliminary step, the transport properties of the bare Ba-block were measured in the Van der Pauw geometry. Both Ba-Cu-O and $(Ba_{0.9}Nd_{0.1})$-Cu-O films, about 200 Å thick, were grown in the same conditions as for the superconducting superlattices. In Fig.1 the resistivity versus temperature behavior is shown. Both samples are not superconducting. However, a major difference can be noticed in the temperature dependence of their resistivity: the Ba-Cu-O film shows a more marked semiconducting behavior while, in the case of the Nd substituted sample, resistivity is almost constant in temperature and its value is about one order of magnitude smaller than in the previous case. This result shows that the main effect of the Nd substitution is a sizeable increase of the metallic character of the film. This issue has been confirmed by thermopower measurements as illustrated in the inset of fig. 1. The Seebeck effect of $(Ba_{0.9}Nd_{0.1})$-Cu-O thin film is positive and in the temperature range 80 – 290 K it has a value typical of common metals of about 1-2 µV/K. Successively the Seebeck effect was investigated in both pure and Nd substituted superlattices.

Firstly we consider $(BaCuO_{2+\delta})_2/(CaCuO_2)_n$ superlattices with different periodicity ($n$=2, 3, 4). In Fig. 2 (upper panel) we report the thermopower of such superlattices as a function of temperature. The whole curve from room temperature down to the transition temperature is shown only for the 2×2 superlattice ($T_C$= 69 K), while for the 2×3 and 2×4 superlattices only measurements above 80 K are presented. The Seebeck effect of these artificial samples qualitatively follows the universal behavior displayed by cuprates, showing the typical characteristics of samples in the optimal or underdoped regimes. In 2×2 superlattices, $S(T)$ behaves like an almost optimally doped sample, being zero in superconducting state, rising towards a maximum just above $T_c$ before decreasing linearly with temperature. As the number of $CuO_2$ planes increases (2×3 and 2×4 samples), $S(T)$ increases in value and the broad maximum shifts towards higher temperatures, as expected for a lower doping per $CuO_2$ plane. In table 1 we summarize the values of Seebeck coefficient at 290 K ($S(290)$) and $p_{pl}$ calculated from eq. (1) for all samples. We find that the 2×2 superlattice is slightly underdoped ($p_{pl}$~0.13), while the doping decreases by increasing the $CuO_2$ planes and for the 2×4 superlattice we estimate $p_{pl}$ ~0.077.

We consider now the Nd doped superlattices. Fig. 2 (lower panel) shows the thermopower measurements on $(Ba_{0.9}Nd_{0.1}CuO_{2+\delta})_2/(CaCuO_2)_n$ superlattices with $n$=1, 2, 6 and 11. A first interesting remark that can be extracted from the data is that the 2×1 superlattice has lower thermopower values than all the other superlattices, indicating a doping level closer to optimal. At the same time the 2×1 superlattice has a lower transition temperature and higher electrical resistivity relative to superlattices with different periodicity: both these features could be explained in terms of underdoping and/or structural disorder. It was pointed out in ref. [17] that the role of the disorder at the interface strongly influences the transport properties and that this effect becomes more and more important as the number of $CuO_2$ planes in the IL block decreases. On the other hand, thermopower, which is less affected than resistivity by the disorder, indicates a larger doping in the 2×1 than in the 2×2 superlattice, clearly attributing the increase of resistivity and the depression of superconductivity in the 2×1 superlattice to structural disorder.

Concerning the thermopower dependence on the number of adjacent $CuO_2$ planes, the overall behavior appears different. Starting from the lowest $n$ value the thermopower increases with the number of $CuO_2$ planes up to $n$=6, but the 2×11 superlattice has lower thermopower than the 2×6 superlattice.

In table 2 we report $S(290)$ and $p_{pl}$ for the $(Ba_{0.9}Nd_{0.1}CuO_{2+\delta})_2/(CaCuO_2)_n$ superlattices calculated from eq. (1). The holes per $CuO_2$ plane calculated from the OCT law show the following trend: for the 2×1 superlattice $p_{pl}$~0.14, but with increasing $n$ the calculated $p_{pl}$ decreases slightly and tends to saturate slightly above 0.1 ($p_{pl}$ =0.105 and 0.106 for *n*=6 and 11 samples respectively). At first sight, this result could suggest that Nd substitution affects the charge transfer mechanism. On the other hand, the critical temperatures of these superlattices are similar to those of superlattices without Nd, indicating that the doping in the $CuO_2$ planes is not strongly modified by the presence of the rare earth [14] [18].

A second possibility is that the Nd doped CR block contributes to the thermopower differently from the undoped CR block. To investigate this issue we have to model the Seebeck effect in a system of parallel layers. The diffusive term of the Seebeck effect is given by the Mott formula $S = \dot{\sigma}/\sigma$ where $\dot{\sigma} = \partial\sigma/\partial\varepsilon\big|_{\varepsilon_F}$ is the derivative of the electrical conductivity $\sigma$ calculated at the Fermi level. For N channels which conduct in parallel, this expression can be generalized in the following way:

$$S = \frac{\dot{\sigma}_{TOT}}{\sigma_{TOT}} = \frac{1}{\sigma_{TOT}}\sum_i \sigma_i \cdot S_i \qquad (2)$$

where $\sigma_{TOT} = \sum_i \sigma_i$, $\dot{\sigma}_{TOT} = \sum \dot{\sigma}_i$ and $S_i = \dot{\sigma}_i/\sigma_i$ is the Seebeck effect of the i-channel.

In the case of superlattices, $\sigma_{TOT}$ is the sum of the conductivity of IL ($\sigma_{IL}$) and CR ($\sigma_{CR}$) blocks, $\sigma_{TOT} = \sigma_{IL} + \sigma_{CR}$. Assuming that the conductivity of CR block is negligible ($\sigma_{CR} << \sigma_{IL}$) and that each $CuO_2$ plane has the same conductivity ($\sigma_{IL} = m \cdot \sigma_{pl}$, where *m* is the number of $CuO_2$ planes per unit cell) from eq. (2) we obtain:

$$S = \sum_i \frac{\sigma_i}{\sigma_{TOT}} S_i = m \frac{\sigma_{pl}}{m \cdot \sigma_{pl}} S_{pl} = S_{pl} \qquad (3)$$

Therefore the aforementioned hypotheses are necessary in order to obtain a thermopower which is related only to the doping of the $CuO_2$ planes, as occurs in the majority of conventional cuprates.

The condition $\sigma_{CR} << \sigma_{IL}$ is certainly valid for $(BaCuO_{2+\delta})_2/(CaCuO_2)_n$ superlattices for intermediate periodicity. Indeed, the resistivity of the Ba-Cu-O film ($\rho \geq 1.5 \cdot 10^{-4}$ $\Omega \cdot m$) is three times larger than those of $(BaCuO_{2+\delta})_2/(CaCuO_2)_n$ superlattices even for *n*~8 [2]. However the case of the $(Ba_{0.9}Nd_{0.1}CuO_{2+\delta})_2/(CaCuO_2)_n$ superlattices could be different, the resistivity of $(Ba_{0.9}Nd_{0.1})$-Cu-O film being much lower ($\rho \sim 1 \cdot 10^{-5}$ $\Omega \cdot m$). Assuming that the CR block transfers always the same charge, independently of the IL block thickness, the charge per plane would be related to the number of $CuO_2$ planes in the IL block by $p_{tot}=m \cdot p_{pl}$ (where *m=n+1* is the number of $CuO_2$ planes in the IL block when the number of $CaCuO_2$ unit cells in the superlattice is *n*). Thus, for a $(BaCuO_{2+\delta})_2/(CaCuO_2)_n$ superlattice the room temperature Seebeck coefficient can be expressed as a function of $p_{pl}=p_{tot}/(n+1)$:

$$S = S_{pl} = S^{OCT}\left(\frac{p_{tot}}{n+1}\right) \quad (4)$$

where the right side is given by eq. (1). In fig. 3 we plot the experimental values $S(290)$, for the $(BaCuO_{2+\delta})_2/(CaCuO_2)_n$ superlattices as a function of the number of $CuO_2$ planes. We can see that the exponential increase of the experimental data is well fitted by eq. (1) with $p_{tot}=0.39\pm0.1$ holes (continuous line). A close inspection of eq.(2)-(3) reveals that we cannot exclude small not uniformity of the charge distribution between the planes. In fact in eq. (2) there is a sum of the products between conductivity and Seebeck effect of each plane: the first decreases with the doping while the latter increases. If the electrical conductivity compensates the changes of $S$ among the planes a not uniformity of the order of 10% cannot be detected. Anyway charge distribution can be inspected by means other observations, such as the width of the superconducting transition; a not uniformity of 10% between the planes modifies the critical temperature of adjacent planes by nearly 30%, yielding a much broader transition than that experimentally observed. On the other hand, regardless the uniformity of charge distribution, we can reliably conclude that all the $(n+1)$ planes have to participate to the conduction and in this condition the total charge transferred by the CR block can be accurately estimated.

Finally, our results support the assumption of a nearly equal distribution of charge among all the $(n+1)$ $CuO_2$ planes in the IL block and gives good evidence that the total charge from the CR block is always the same. Extrapolating this result for $n$ lower than 2 and larger than 4, we expect that the optimal doping $p_{opt} \sim 0.16$ is obtained with $p_{tot}/(n+1) \sim 0.16$, i.e. $2 < n+1 < 3$ and that superconductivity disappears in superlattices with $p_{tot}/(n+1) < p_{min} \sim 0.04$, i.e. $n+1 > 10$. The behavior of the critical temperature of the $(BaCuO_{2+\delta})_2/(CaCuO_2)_n$ superlattices as a function of $CuO_2$ planes [17] agrees well with these expectations. These results show that Seebeck effect in $(BaCuO_{2+\delta})_2/(CaCuO_2)_n$ superlattices follows closely the universal behavior of cuprates and in such a case the OCT law is an useful tool to investigate the charge transfer mechanism.

For the $(Ba_{0.9}Nd_{0.1}CuO_{2+\delta})_2/(CaCuO_2)_n$ superlattices, the analysis is less trivial because of the circumstance that the Nd doped CR block is not insulating and that, consequently, the hypothesis $\sigma_{CR} << \sigma_{IL}$ fails. In this case eq. (2) can be expressed as:

$$S = \sum_i \frac{\sigma_i}{\sigma_{TOT}} S_i = \frac{(m \cdot \sigma_{pl} \cdot S_{pl} + \sigma_{CR} \cdot S_{CR})}{m \cdot \sigma_{pl} + \sigma_{CR}} = \frac{(\sigma_{IL} \cdot S_{pl} + \sigma_{CR} \cdot S_{CR})}{\sigma_{IL} + \sigma_{CR}} \quad (5)$$

In order to make use of eq. (5), we have to make the reasonable assumption that the results found for the $(BaCuO_{2+\delta})_2/(CaCuO_2)_n$ superlattices for $n$ up to 4 (namely that the charge transferred from the CR block is the same for all the superlattices and it is equally distributed among the $m=n+1$

CuO$_2$ planes) are also valid for Nd doped superlattices and for $n$ up to 11. In this framework, the room temperature Seebeck coefficient becomes:

$$S = \frac{\sigma_{IL}(n+1) \cdot S^{OCT}(p_{tot}/(n+1)) + \sigma_{CR} \cdot S_{CR}}{\sigma_{IL}(n+1) + \sigma_{CR}} \quad (6)$$

The room temperature Seebeck coefficient depends on several parameters: $\sigma_{IL}(n+1)$, $p_{tot}$, $\sigma_{CR}$ and $S_{CR}$. To understand the effect of a conducting CR block we choose reasonable values for these parameters: we fix $p_{tot}=0.39$ holes, previously estimated for Nd-free superlattices; for $S_{CR}$ and $\sigma_{CR}$ we use the experimental values for the bare Ba$_{0.9}$Nd$_{0.1}$CuO$_{2+\delta}$ block ($S_{CR}=1.5$ µV/K and $\sigma_{CR}=10^5$ ($\Omega \cdot$m)$^{-1}$), assuming that the metallic properties of the CR block are only slightly affected by the charge transfer towards the IL block. For $\sigma_{IL}$ we assume the simple form $\sigma_{IL}(n+1) = 4 \cdot 10^6 \cdot (n+1)^{-2.5}$ ($\Omega \cdot$m)$^{-1}$ which best fits the resistivity behavior of superlattices with different periodicities [2]. The result is plotted in fig. (3) (dotted line); this curve increases up to $n+1 \approx 7$-8 and then slowly decreases, showing the same behavior of the experimental $S(290)$ values of the (Ba$_{0.9}$Nd$_{0.1}$CuO$_{2+\delta}$)$_2$/(CaCuO$_2$)$_n$ superlattices. The fitting of experimental data can be easily improved by slightly changing the parameters of eq. (6), but there is little point in doing this, due to the large number of free parameters and to the complexity of the mechanisms into play. Our goal is to account qualitatively for the decrease of the thermopower at large $n$ values, as a consequence of the presence of a metallic CR block. Indeed, for thin IL block, $\sigma_{CR}$ can be considered still much smaller than $\sigma_{IL}$ leading to an increase of $S(290)$ as $p_{pl}$ decreases. However, but for large $n$ values, the conductivity of the CR block becomes dominant and the measured thermopower approaches $S_{CR}$.

Summarizing, the OCT law is valid when the CR block conductivity is negligible compared with that of IL block. On the other hand, when the CR block is metallic and the IL block is thick, the parallel layer conduction must be considered.

**Conclusions**

Thermoelectric power has been investigated in (BaCuO$_{2+\delta}$)$_2$/(CaCuO$_2$)$_n$ and (Ba$_{0.9}$Nd$_{0.1}$CuO$_{2+\delta}$)$_2$ /(CaCuO$_2$)$_n$ superconducting superlattices. Even though the analysis, in the case of the undoped artificial structure, is restricted to $n=4$, we believe that the large difference in the behavior of $S$ versus $n$ for Nd doped and undoped superlattices is a clear evidence for the proposed model. It was found that the different degree of metallic behavior of the Ba-block strongly influences the overall Seebeck effect of the superlattices. A simple model, which assumes parallel Ba-Cu-O and Ca-Cu-O blocks, was considered to take into account such an effect. Predictions based on this model were found in good agreement with experimental results. It was found that charge carriers (holes) transferred from the CR block are equally distributed among the $CuO_2$ planes in the IL block. Consequently the decrease of $p_{pl}$, for thicker IL blocks, gives rise to a metal-insulator transition for *n>11*. The *n=1* superlattices represent an interesting case. This superlattice, according to the Seebeck effect measurements, results to be close to optimally doped. However, its transition temperature is much lower than the *n=2* sample. Seebeck effect measurements demonstrate that the worsening of the superconducting properties of this superlattice can be ascribed to interface disorder rather than to the decrease of carrier concentration in the IL block. Finally, it is worth to notice that Nd substitution does not change the doping capability of the CR block, while strongly varying its metallic character. This can open interesting perspectives for the realization of superlattices with more metallic CR blocks and consequently lower anisotropy and higher critical current density.

**FIGURE CAPTIONS**

**Figure 1**. Resistivity as a function of temperature of Ba-Cu-O and $(Ba_{0.9}Nd_{0.1})$-Cu-O films. In the inset the Seebeck coefficient of the $(Ba_{0.9}Nd_{0.1})$-Cu-O film is reported.

**Figure 2**. Upper panel: Seebeck effect of $(BaCuO_{2+\delta})_2/(CaCuO_2)_n$ super lattices with $n=2,3,4$. Lower panel: Seebeck effect of $(Ba_{0.9}Nd_{0.1}CuO_{2+\delta})_2/(CaCuO_2)_n$ super lattices with $n=1, 2, 6, 11$.

**Figure 3**. $S(290)$ versus the number of $CuO_2$ planes for the $(BaCuO_{2+\delta})_2/(CaCuO_2)_n$ superlattices with $n=2, 3, 4$ and $(Ba_{0.9}Nd_{0.1}CuO_{2+\delta})_2/(CaCuO_2)_n$ superlattices with $n=1, 2, 3, 6, 11$. Continuous line is obtained from eq. (4) with $p_{tot}=0.39$. The dotted line is obtained from eq. (4) with $p_{tot}=0.39$, $S_{CR}=1.5$ μV/cm, $\sigma_{CR}=10^5$ $(\Omega \cdot m)^{-1}$ and $\sigma_{IL}(n+1) = 4 \cdot 10^6 \cdot (n+1)^{-2.5}$ $(\Omega \cdot m)^{-1}$.

**TABLE CAPTIONS**

**Table I.** $(BaCuO_{2+\delta})_2/(CaCuO_2)_n$ superlattices: room temperature Seebeck coefficient $S(290)$; the holes per $CuO_2$ plane $p_{pl}$ calculated by eq. (1) [16]; $(n+1) \cdot p_{pl}$ which represents the total charge per supercell, $p_{tot}$, in the hypothesis of equal distribution of charge among the $CuO_2$ planes.

**Table II.** $(Ba_{0.9}Nd_{0.1}CuO_{2+\delta})_2/(CaCuO_2)_n$ superlattices: room temperature Seebeck coefficient $S(290)$; the holes per $CuO_2$ plane $p_{pl}$ calculated by eq. (1) [16]. In this case, due to the questionable use of the OCT law, $p_{pl}$ values loose a straightforward meaning while a more reliable estimation is given by $p_{tot}/(n+1)$ with $p_{tot}=0.39$ (right column).

**Table I.**

| 2×n | $S(290)$, μV/K | $p_{pl}$ | $(n+1)\cdot p_{pl}$ |
|---|---|---|---|
| 2×2 | 11 | 0.13 | 0.39 |
| 2×3 | 32 | 0.10 | 0.40 |
| 2×4 | 78 | 0.077 | 0.39 |

**Table II**

| 2×n | $S(290)$, μV/K | $p_{pl}$ | $p_{tot}/(n+1)$ |
|---|---|---|---|
| 2×1 | 6.6 | 0.14 | 0.19 |
| 2×2 | 13.5 | 0.12 | 0.13 |
| 2×3 | 19 | 0.114 | 0.1 |
| 2×6 | 27 | 0.105 | 0.06 |
| 2×11 | 26 | 0.106 | 0.03 |

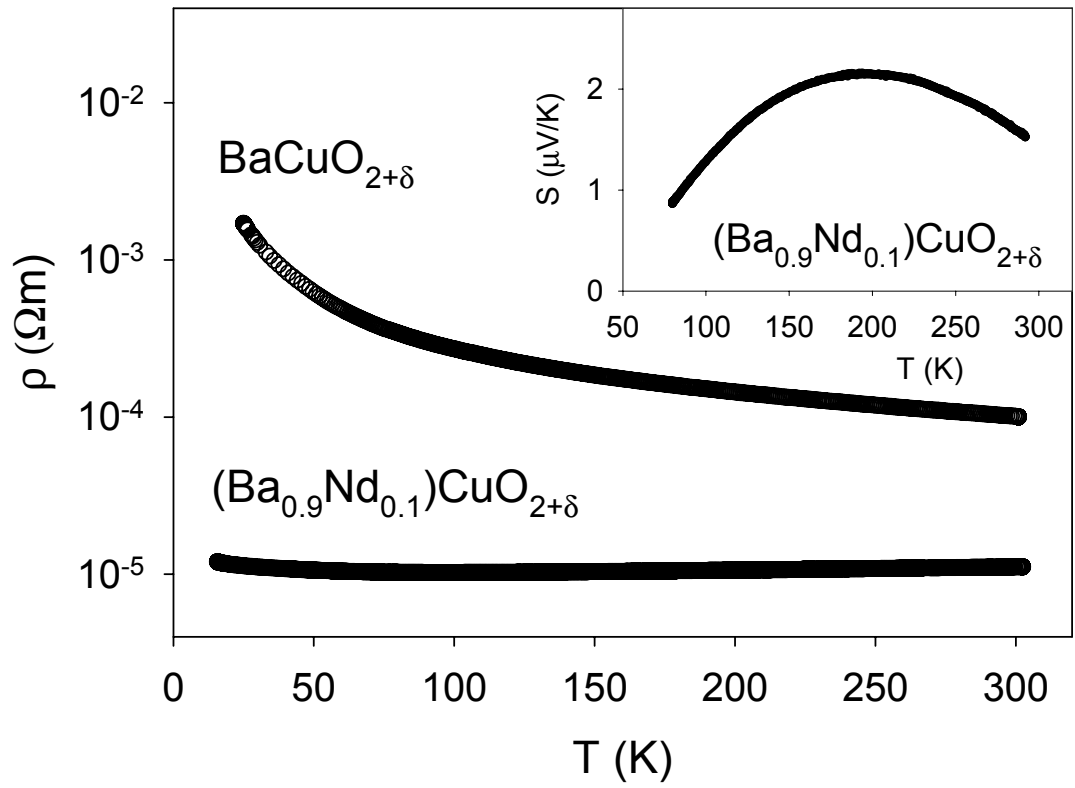

Figure 1

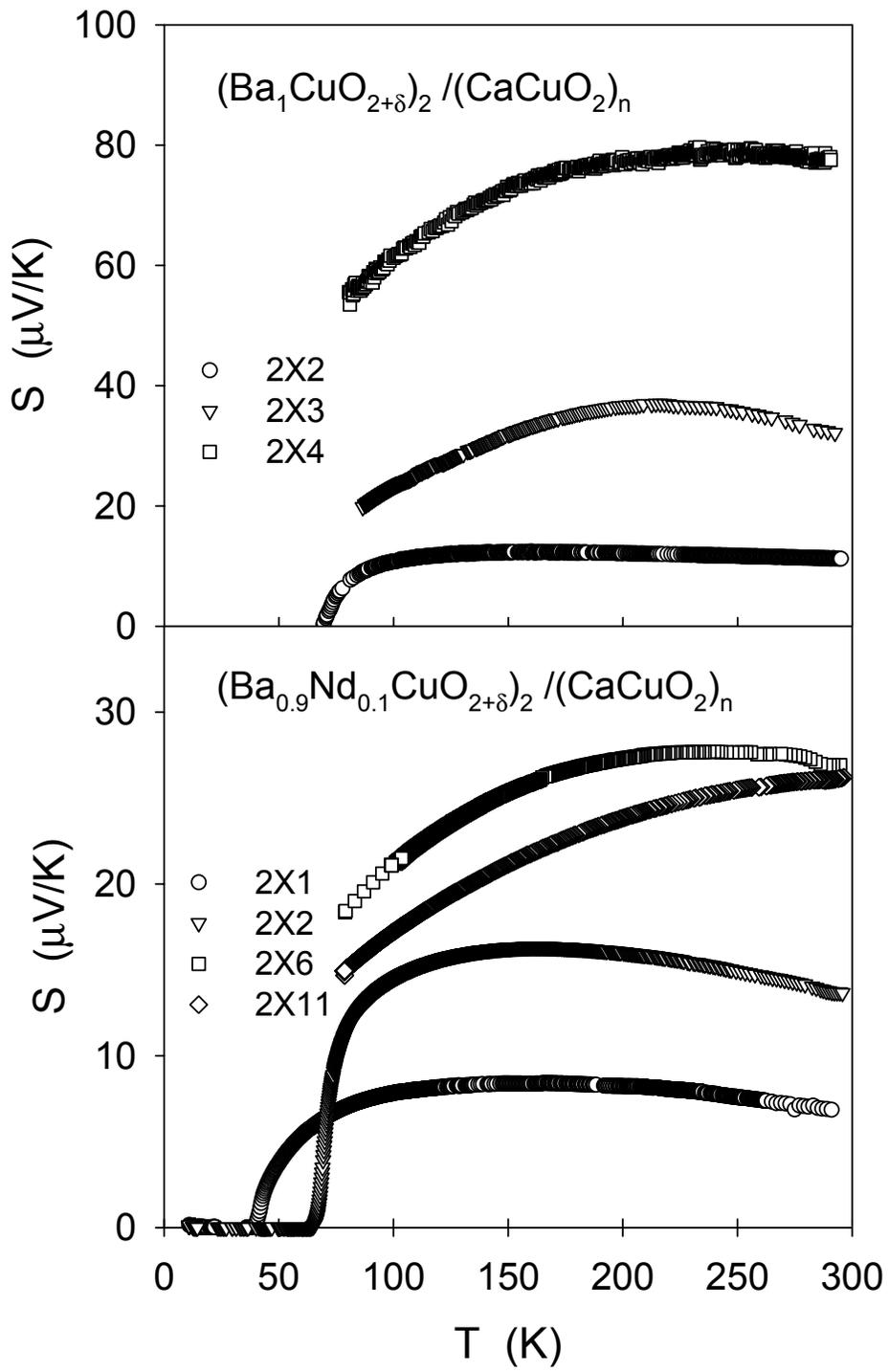

Figure 2

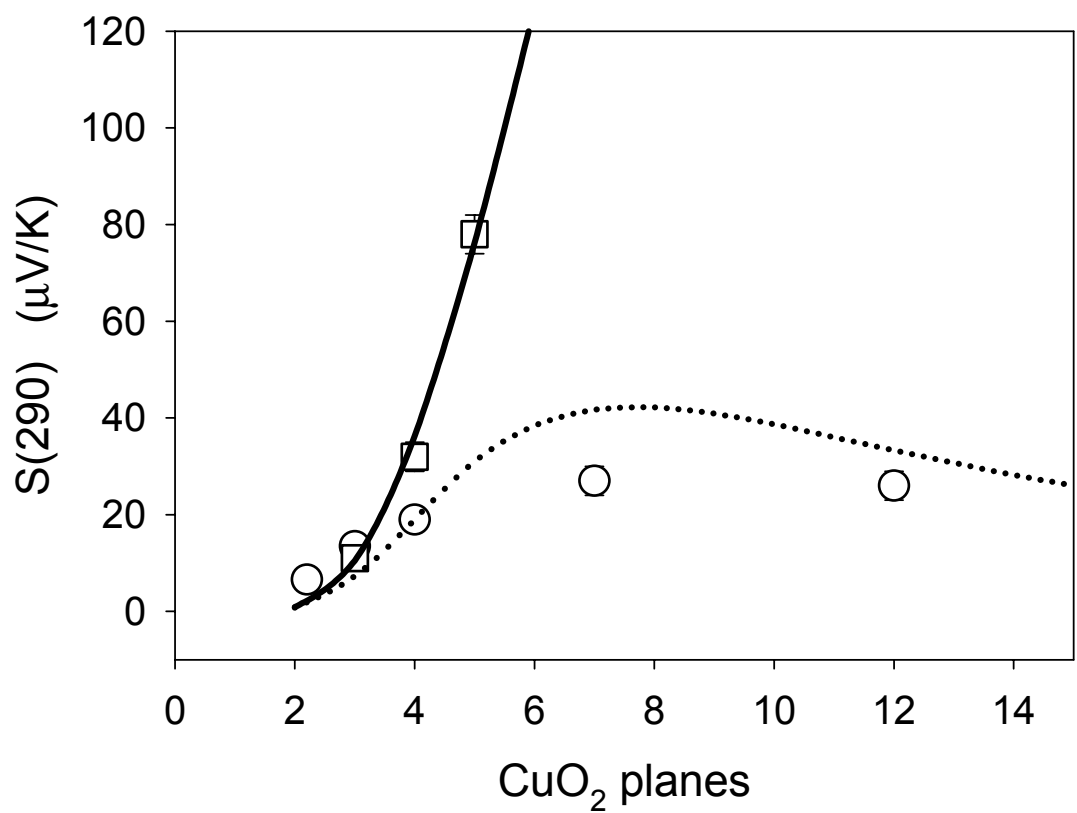

Figure 3